\documentclass[cite,a4paper,12pt]{article}
\usepackage{amsmath}
\usepackage{myart}
\usepackage{graphicx}

\usepackage{amsfonts}
\usepackage{amssymb}

\usepackage[cp1251]{inputenc}
\usepackage[english,russian]{babel}

\usepackage{cite}
\usepackage{myart}

\begin{document}
\hfill\hbox{PTA/06-05}

\hfill\hbox{April 2006}

\selectlanguage{russian}
\parbox{9cm}
{``Физик-теоретик -- это ответственный работник. Небрежные и
неправильные теоретические ста\-тьи приносят науке не пользу, а вред.
Как под\-тасовки в эксперименте \cite{KAT}.''
\vskip 0.4cm
К.А.Тер-Мартиросян}

\fboxrule=0.05in
\fboxsep=0.0in
\vskip -0.9in
\hfill%
\parbox[t]{3.5cm}{%
\hspace{0.15cm}\fbox{%
\parbox[b][1.1\height]{0.98in}{%
\includegraphics[
height=1.2886in,
width=1.0646in
]%
{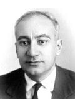}%
\vskip 0.0cm
}
}
\vskip 0.3cm
Памяти Учителя%
}%

\bigskip
\selectlanguage{english}

\begin{center}
{\Large \textbf{Gravitational Yang-Lee Model.}}

{\Large \textbf{Four Point Function}}

\vspace{1.2cm}

{\large Al.Zamolodchikov}\footnote{On leave of absence from Institute of
Theretical and Experimental Physics, B.Cheremushkinskaya 25, 117259 Moscow,
Russia.}

\vspace{0.2cm}

Laboratoire de Physique Math\'ematique\footnote{Laboratoire Associ\'e au CNRS
UMR 5825}

Universit\'e Montpellier II

Pl.E.Bataillon, 34095 Montpellier, France

\vspace{0.2cm}

and

\vspace{0.2cm}

Service de Physique Th\'eorique

CNRS - URA 2306, C.E.A. - Saclay

F-91191, Gif-sur-Yvette, France
\end{center}

\textbf{Abstract}

The four-point perturbative contribution to the spherical partition
function of the gravitational Yang-Lee model is evaluated
numerically. An effective integration procedure is due to a
convenient elliptic parameterization of the moduli space. At certain
values of the ``spectator'' parameter the Liouville four-point
function involves a number of ``discrete terms'' which have to be
taken into account separately. The classical limit, where only
discrete terms contribute, is also discussed. In addition, we
conjecture an explicit expression for this partition function at the
``second solvable point'' where the spectator matter is in fact
another $\mathcal{M}_{2/5}$ (Yang-Lee) minimal model.

\section{Introduction}

This work is a direct continuation of the previous developments of
refs.\cite{Classic} and \cite{GYL} where a kind of analytic-numeric
extrapolation of the perturbative series has been applied to study
the spherical partition function of a perturbed conformal matter
immersed to the quantized 2D gravity. In these works the scaling
Yang-Lee model (complemented with a conformal ``spectator matter'')
is taken as an example of the perturbed matter theory. Similar
development for the massive free Majorana fermions (the scaling
Ising model) coupled to 2D gravity is reported in ref. \cite{Yuki}.

For the further convenience we summarize in this section few main
concepts of \cite{GYL}. This is mostly to facilitate subsequent
references and introduce notations. It doesn't make the present
report self consistent, a minimum acquaintance with the previous one
\cite{GYL} is assumed.

\textbf{Gravitational Yang-Lee} model (GYL) can be conventionally
defined through the Lagrangian density
\begin{equation}
\mathcal{L}_{\text{GYL}}=\mathcal{L}_{\text{matter}}+\mathcal{L}_{\text{L}%
}+\mathcal{L}_{\text{gh}}\label{GYL}%
\end{equation}
where $\mathcal{L}_{\text{L}}$ is referred to as the Liouville
component of the theory
\begin{equation}
\mathcal{L}_{\text{L}}=\frac1{4\pi}(\partial_{a}\phi)^{2}+\mu
e^{2b\phi
}\label{LL}%
\end{equation}
and $\mathcal{L}_{\text{gh}}$ is the Lagrangian of the
$c_{\text{gh}}=-26$ conformal ghost $BC$ system. As usual, $\phi$ is
the Liouville field, $\mu$ is the cosmological constant and $b$ is a
parameter related to the Liouville central charge
$c_{\text{L}}=1+6Q^{2}$ through the ``background charge''
$Q=b+b^{-1}$. The matter component consists of some unperturbed
``spectator'' CFT with central charge $c_{\text{sp}}$ and the
$\mathcal{M}_{2/5}$ minimal CFT model (critical Yang-Lee model
\cite{Cardy} with $c_{\text{YL}}=-22/5$) perturbed by the only
$\mathcal{M}_{2/5}$ non-trivial primary field $\varphi$, i.e., the
basic Yang-Lee field of dimension $\Delta=-1/5$
\begin{equation}
\mathcal{L}_{\text{matter}}=\mathcal{L}_{\text{sp}}+\mathcal{L}_{\text{YL}%
}+\frac{i\lambda}{2\pi}\varphi(x)e^{2g\phi}\label{Lmatter}%
\end{equation}
Parameters $b$ in (\ref{LL}) and $g$ in the interaction term of
eq.(\ref{Lmatter}), are determined by $c_{\text{sp}}$ and $\Delta$
through the balance equations
\begin{align}
c_{\text{sp}}+c_{\text{YL}}+c_{\text{L}}+c_{\text{gh}}  & =0\label{balance}\\
g(Q-g)+\Delta & =1\nonumber
\end{align}
In the case of spherical geometry, which we only consider in the
present study, any details of the spectator matter are not important
except for the parameter $c_{\text{sp}}$. This component is added to
get a formal access to the parameter $b^{2}$ of the model, in
particular, to have a link with the classical limit. The interaction
term contains a dimensional coupling constant
$\lambda\sim\mu^{1/\rho}$ where
\begin{equation}
\rho=bg^{-1}\label{rho}%
\end{equation}

\textbf{The spherical partition function} $Z(\mu,\lambda)$ of the
model is developed as a systematic perturbative series in the
coupling constant
$\lambda$%
\begin{equation}
\frac{Z(\mu,\lambda)}{Z(\mu,0)}=\sum_{n=0}^{\infty}\frac{(-i\lambda)^{n}%
}{(2\pi)^{n}n!}\left\langle \left\langle \left(  \varphi
e^{2g\phi}\right)
^{n}\right\rangle \right\rangle _{\text{LG}}\label{Zexp}%
\end{equation}
where $\left\langle \left\langle \left(  \varphi e^{2g\phi}\right)
^{n}\right\rangle \right\rangle _{\text{LG}}$ are (integrated and
normalized) $n$-point functions in the Liouville gravity (i.e., at
$\lambda=0$).

The normalized correlation functions scale as\footnote{Here
notations differ slightly from those of \cite{GYL}, where the
coefficinets $a_{n}$ were dimensional and included the multiplier
$(2\pi)^{n}n!$.}
\begin{equation}
\left\langle \left\langle \left(  \varphi e^{2g\phi}\right)
^{n}\right\rangle
\right\rangle _{\text{LG}}=(\pi\mu)^{-n/\rho}a_{n}\label{an}%
\end{equation}
with some dimensionless numbers $a_{n}$, i.e., the perturbative
development is in fact a series in powers of the dimensionless
scaling parameter $\lambda \mu^{-1/\rho}$. It is convenient to
introduce also the unnormalized correlation functions
\begin{equation}
\left\langle \left(  \varphi e^{2g\phi}\right)  ^{n}\right\rangle _{\text{LG}%
}=(\pi\mu)^{Q/b-n/\rho}G_{n}\label{Gn}%
\end{equation}
In the Liouville gravity these correlation functions are evaluated
as the integrals of the products of the Liouville and
$\mathcal{M}_{2/5}$ $n$-point functions
\begin{align}
\left\langle \left(  \varphi e^{2g\phi}\right)  ^{n}\right\rangle
_{\text{LG}}  & =\left\langle C\bar C(x_{1})C\bar C(x_{2})C\bar C(x_{3}%
)\right\rangle _{\text{gh}}\times\label{Mn}\\
& \int\left\langle \varphi(x_{1})\ldots\varphi(x_{n})\right\rangle
_{\text{YL}}\left\langle V_{g}(x_{1})\ldots
V_{g}(x_{n})\right\rangle _{\text{L}}d^{2}x_{4}\ldots
d^{2}x_{n}\nonumber
\end{align}
where $\left\langle \ldots\right\rangle _{\text{gh}}$, $\left\langle
\ldots\right\rangle _{\text{YL}}$ and $\left\langle
\ldots\right\rangle _{\text{L}}$ are related to respectively ghost,
$\mathcal{M}_{2/5}$ minimal
model and (unnormalized) Liouville correlations. We also denote $V_{g}%
=\exp(2g\phi)$. Geometrically the $n-3$ dimensional integral in
(\ref{Mn}) is the integral over the moduli space of a sphere with
$n$ punctures.

As in \cite{GYL}, in this paper we are mainly interested in the
fixed area partition function $Z_{A}(\lambda)$, which is related to
$Z(\mu,\lambda)$ as
\begin{equation}
Z(\mu,\lambda)=\int_{(0)}^{\infty}Z_{A}(\lambda)e^{-\mu A}\frac{dA}A\label{ZA}%
\end{equation}
The lower limit $(0)$ here is simply a particular prescription how
to regularize the divergency of the integral at small $A$
\cite{GYL}. The fixed area partition function has the following
scaling form
\begin{equation}
Z_{A}(\lambda)=Z_{A}(0)\,z(h)\label{ZA0}%
\end{equation}
where $Z_{A}(0)$ scales as $A^{-Q/b}$. The scaling function $z(h)$
is a regular expansion
\begin{equation}
z(h)=\sum_{n=0}^{\infty}z_{n}(-h)^{n}\label{zh}%
\end{equation}
in the fixed area dimensionless scaling parameter
\begin{equation}
h=\lambda\left(  \frac A\pi\right)  ^{1/\rho}\label{h}%
\end{equation}
Apparently, coefficients $z_{n}$ are related to the numbers $G_{n}$
in (\ref{Gn}) as
\begin{equation}
z_{n}=\frac{a_{n}\Gamma(-1-b^{-2})}{(2\pi)^{n}n!\Gamma(n\rho^{-1}-b^{-2}%
-1)}\label{zn}%
\end{equation}
Sometimes it is convenient to express these coefficients directly
\begin{equation}
z_{n}=\frac{\mathcal{I}_{n}}{(2\pi)^{n}n!Z_{\text{L}}^{(A)}}\label{znA}%
\end{equation}
through the non-normalized fixed area correlation functions
\begin{equation}
\left\langle \left(  \varphi e^{2g\phi}\right)  ^{n}\right\rangle
^{(A)}=\left(  \frac A\pi\right)
^{n/\rho-Q/b}\mathcal{I}_{n}=\frac{\left(
A\pi^{-1}\right)  ^{n/\rho-Q/b}G_{n}}{\Gamma(ngb^{-1}-1-b^{-2})}\label{GA}%
\end{equation}
and the fixed area Liouville partition function
\begin{equation}
Z_{\text{L}}^{(A)}=\frac{\gamma^{1/b^{2}}(b^{2})}{\pi^{3}b\Gamma(b^{-2}%
-1)}\label{ZL}%
\end{equation}
It was argued in \cite{GYL} that $z(h)$ is an entire function of
$h$.

\textbf{The infinite volume specific energy} $\mathcal{E}_{\text{vac}}%
(b^{2})=-\mu_{\text{c}}$ enters the leading $A\rightarrow\infty$
asymptotic of
$Z_{A}(\lambda)$%
\begin{equation}
Z_{A}(\lambda)\sim Z_{\infty}\left(  \frac A\pi\right)  ^{-Q_{\text{IR}%
}/b_{\text{IR}}}\exp\left(  -\mathcal{E}_{\text{vac}}(b)A\right) \label{Evac}%
\end{equation}
This quantity scales as
\begin{equation}
\mu_{\text{c}}=f_{0}(b^{2})\lambda^{\rho}\label{f0}%
\end{equation}
Here we introduced dimensionless function $f_{0}(b^{2})$ which is an
important universal characteristic of GYL. Numerical study of this
quantity at different values of the ``spectator'' parameter $b^{2}$
is the main topic of ref.
\cite{GYL} and of the present article. Parameters $Q_{\text{IR}}=b_{\text{IR}%
}^{-1}+b_{\text{IR}}$ and $b_{\text{IR}}$ in (\ref{Evac}) are fixed
by the ``IR central charge balance'' \cite{GYL}
\begin{equation}
1+6Q_{\text{IR}}^{2}+c_{\text{sp}}=26\label{IRbalance}%
\end{equation}

In terms of the scaling function $z(h)$ eq.(\ref{Evac}) reads as the
following asymptotic behavior at $h\rightarrow\infty$
\begin{equation}
\log z(h)=\pi f_{0}(b)h^{\rho}+\left(  \delta+1/2\right)  \log
h+O(1)\label{logzh}%
\end{equation}
where
\begin{equation}
\delta+1/2=\rho\left(
Qb^{-1}-Q_{\text{IR}}b_{\text{IR}}^{-1}\right)
\label{delta}%
\end{equation}

All the above considerations are either kinematical or based on
natural physical assumptions. A less trivial observation is that in
GYL (as well as in a number of other important gravitational models,
see e.g., \cite{Yuki}) the asymptotic (\ref{logzh}) holds in the
whole complex plane away from the negative real axis and that all
zeros of this entire function $z(h)$ are real. This property has
been established numerically in the classical limit
$b^{2}=0$ \cite{Classic} and in the special, exactly solvable case $b^{2}%
=0.4$. It is then extended, as a conjecture, to the whole region of
the parameter. Combined with the asymptotic behavior (\ref{logzh})
this feature leads to an effective analytic-numeric algorithm, which
allows to restore the scaling function to an impressive accuracy
starting from a few first perturbative coefficients $z_{n}$ in the
expansion (\ref{zh}) (see \cite{GYL} for more details).

In the previous study we used these coefficients up to $z_{3}$,
where the numerical evaluation of the matter and Liouville
correlation functions doesn't offer any technical difficulties. In
the present article we develop a method of numerical integration
over moduli, which allows to calculate the four-point function in
(\ref{Mn}) to a precision sufficient to improve the results of
\cite{GYL}.

For completeness we quote here the expressions for $z_{2}$ and
$z_{3}$ used in \cite{GYL}
\begin{align}
z_{2}  & =-\frac{\gamma^{1-2g/b}(b^{2})\Gamma(b^{-2}-1)\gamma(2gb-b^{2}%
)}{8\Gamma(1+b^{-2}-2gb^{-1})}\label{z2z3}\\
z_{3}  & =\kappa\frac{\gamma^{1-3g/b}(b^{2})\Gamma(b^{-2}-1)\gamma
(3gb-b^{2})\Upsilon_{b}(b)\Upsilon_{b}^{3}(2g)}{48b\Gamma(2+b^{-2}%
-3gb^{-1})\Upsilon_{b}(3g)\Upsilon_{b}^{3}(g)}\nonumber
\end{align}
where
\begin{equation}
\kappa=\frac{\gamma^{3/2}(1/5)}{5\gamma^{1/2}(3/5)}=1.91131\ldots\label{kappa}%
\end{equation}
in the last expression is related to the basic structure constant
$C_{\varphi\varphi\varphi}$ in the critical Yang-Lee model
\cite{Cardy}. Special function $\Upsilon_{b}(x)$ is the standard
element of the Liouville field theory construction (see \cite{AAl}
or \cite{threep} for the definitions and properties).

\section{Matter four point function}

In the four-point function the integral (\ref{Mn}) reduces to
\begin{equation}
\left\langle \left(  \varphi e^{2g\phi}\right)  ^{4}\right\rangle _{\text{LG}%
}=\int G_{\text{YL}}(x,\bar x)G_{\text{L}}(x,\bar x)d^{2}x\label{M4}%
\end{equation}
where both in the Yang-Lee four point function
\begin{equation}
G_{\text{YL}}(x,\bar x)=\left\langle \varphi(0)\varphi(1)\varphi
(\infty)\varphi(x)\right\rangle _{\text{YL}}\label{GYL4}%
\end{equation}
and the Liouville one
\begin{equation}
G_{\text{L}}(x,\bar x)=\left\langle V_{g}(0)V_{g}(1)V_{g}(\infty
)V_{g}(x)\right\rangle _{\text{L}}\label{G4L}%
\end{equation}
we have used the projective invariance of the integrand to put
$x_{1}$, $x_{2}$ and $x_{3}$ to $0$, $1$ and $\infty$ respectively.

The $\mathcal{M}_{2/5}$ four point function reads explicitly
\cite{Cardy} as
\begin{equation}
G_{\text{YL}}(x,\bar x)=\mathcal{F}_{I}(x)\mathcal{F}_{I}(\bar
x)-\kappa
^{2}\mathcal{F}_{\varphi}(x)\mathcal{F}_{\varphi}(\bar x)\label{G4YL}%
\end{equation}
where $\kappa$ is from (\ref{kappa}) and the blocks
$\mathcal{F}_{I,\varphi }(x)$ are expressed through the
hypergeometric functions
\begin{align}
\mathcal{F}_{I}(x)  & =x^{2/5}(1-x)^{1/5}{}_{2}F_{1}%
(2/5,3/5,6/5,x)\label{FIphi}\\
\mathcal{F}_{\varphi}(x)  & =x^{1/5}(1-x)^{1/5}{}_{2}F_{1}%
(1/5,2/5,4/5,x)\nonumber
\end{align}

\section{Liouville four point function}

The non-normalized Liouville four point function (\ref{G4L}) can be
represented as an integral over the intermediate momentum $P$
\cite{AAl} (see also \cite{GYL} from where the expressions below are
read off). In our context it looks more convenient to start with the
fixed area $A$ Liouville four-point function
\begin{equation}
G_{\text{L}}^{(A)}(x,\bar x)=\left(  \frac A\pi\right)  ^{(4g-Q)/b}%
g_{\text{L}}(x,\bar x)\label{gL}%
\end{equation}
Function $g_{\text{L}}(x,\bar x)$ in general can be evaluated
through the following integral representation
\begin{equation}
g_{\text{L}}(x,\bar x)=\frac{\mathcal{R}_{g}}{\Gamma(4gb^{-1}-b^{-2}-1)}%
\int^{\prime}\frac{dP}{4\pi}r_{g}(P)\mathcal{F}_{P}\left(  x\right)
\mathcal{F}_{P}\left(  \bar x\right) \label{gG}%
\end{equation}
The prime near the integral sign denotes possible discrete terms
(see below) and
\begin{equation}
\mathcal{F}_{P}\left(  x\right)  =\mathcal{F}_{P}\left(  \left.
\begin{array}
[c]{cc}%
\Delta_{g} & \Delta_{g}\\
\Delta_{g} & \Delta_{g}%
\end{array}
\right|  x\right) \label{Lblock}%
\end{equation}
is the general four point conformal block with all four external
dimensions $\Delta_{g}=g(Q-g)=6/5$, the central charge
$c_{\text{L}}$ and the intermediate dimension $Q^{2}/4+P^{2}.$ This
function was introduced in \cite{BPZ}. In Appendix A we recapitulate
some details end explicit constructions concerning this object. As
in the previous paper \cite{GYL} we use the notations
\begin{equation}
\mathcal{R}_{g}=\left(  \gamma(b^{2})b^{2-2b^{2}}\right)  ^{(Q-4g)/b}%
\frac{\Upsilon_{b}^{4}(b)\Upsilon_{b}^{4}(2g)}{\pi^{2}\Upsilon_{b}%
^{4}(2g-Q/2)}\label{Rg}%
\end{equation}
and
\begin{equation}
r_{g}(P)=\frac{\pi^{2}\Upsilon_{b}(2iP)\Upsilon_{b}(-2iP)\Upsilon_{b}%
^{4}(2g-Q/2)}{\Upsilon_{b}^{2}(b)\Upsilon_{b}^{2}(2g-Q/2-iP)\Upsilon_{b}%
^{2}(2g-Q/2+iP)\Upsilon_{b}^{2}(Q/2-iP)\Upsilon_{b}^{2}(Q/2+iP)}\label{rg}%
\end{equation}
The last function enters the integrand in $P$ in (\ref{gG}) and
therefore a quick numerical evaluation is important. Expression
(\ref{rg}) admits the following integral representation \cite{GYL}
\begin{align}
r_{g}(P)  & =\sinh(2\pi b^{-1}P)\sinh(2\pi bP)\label{rgint}\\
& \ \ \ \times\exp\left(
-8\int_{0}^{\infty}\frac{dt}t\frac{\sin^{2}Pt\left(
\cosh^{2}(Q-2g)t-e^{-Qt}\cos^{2}Pt\right)  }{\sinh bt\sinh
b^{-1}t}\right) \nonumber
\end{align}
(convergent at $g>Q/4$). The integral here is convergent if $g>Q/4$,
i.e., in
our example, at $b^{2}>b_{0}^{2}$ (where $b_{0}^{2}=(11-4\sqrt{6}%
)/5=0.2404\ldots$, see below). At smaller values of $b^{2}$ a
slightly more
complicated expression applies%

\begin{align}
r_{g}(P) &  =\frac{\sinh(2\pi b^{-1}P)\sinh(2\pi bP)\;\gamma^{4}%
(3/2+b^{-2}/2-2gb^{-1})}{\gamma^{2}(3/2+b^{-2}/2-2gb^{-1}+ib^{-1}P)\gamma
^{2}(3/2-b^{-2}/2-2gb^{-1}-ib^{-1}P)}\label{rint}\\
&  \times\exp\left(
-8\int_{0}^{\infty}\frac{dt}t\frac{\sin^{2}Pt\left(
\cosh^{2}(b-2g)t-e^{-Qt}\cos^{2}Pt\right)  }{\sinh bt\sinh
b^{-1}t}\right) \nonumber
\end{align}
This integral representation converges in the interval
$0<b^{2}<(23-4\sqrt {19})/15=0.371\ldots$, which complements the
region of convergence of (\ref{rgint}) (and has essential overlap
with it).

The integral in (\ref{gG}) is understood literally, i.e., it goes
along the real axis, only if $b^{2}>b_{0}^{2}=0.2404\ldots$. At
$b^{2}=b_{0}^{2}$ two double poles of $r_{g}(P)$ at $iP=\pm(Q/2-2g)$
cross the integration contour
and must be picked up explicitly as the discrete terms. Then, at $b^{2}%
=b_{1}^{2}=(4\sqrt{139}-43)/25\approx0.1664\ldots$ the same happens
with the poles at $iP=\pm(Q/2-2g-b)$ and so on. In general the pair
of double poles $iP=\pm(Q/2-2g-nb)$ with $n=0,1,2,\ldots$ shows up
in the form of the discrete term at $b^{2}<b_{n}^{2}$ where
\begin{equation}
b_{n}^{2}=\frac{4\sqrt{25n^{2}+60n+54}-10n-33}{5(4n^{2}+4n-3)}\label{bn}%
\end{equation}
Below we will use the notation
\begin{equation}
P_{n}=i(2g+nb-Q/2)\label{Pn}%
\end{equation}

In the presence of the discrete terms expression (\ref{gG}) reads
\begin{equation}
g_{\text{L}}(x,\bar x)=\sum_{n=0}^{N_{\text{d}}-1}D_{n}(x,\bar
x)+\frac
{\mathcal{R}_{g}}{\Gamma(4gb^{-1}-b^{-2}-1)}\int\frac{dP}{4\pi}r_{g}%
(P)\mathcal{F}_{P}\left(  x\right)  \mathcal{F}_{P}\left(  \bar
x\right)
\label{DndP}%
\end{equation}
where $N_{\text{d}}$ is the actual number of discrete terms
\begin{equation}
N_{\text{d}}=\mathtt{Floor}\left[  \sqrt{1+b^{-4}-14b^{-2}/5}-b^{-2}%
/2+1/2\right] \label{Nd}%
\end{equation}
while in the last ``integral'' term the ordinary integration over
real $P$ is implied. The discrete terms are in a sense
``logarithmic''
\begin{equation}
D_{n}(x,\bar x)=\mathcal{N}_{n}\mathcal{F}_{P_{n}}\left(  x\right)
\mathcal{F}_{P_{n}}\left(  \bar x\right)  \left(  2\operatorname*{Re}f_{P_{n}%
}(x)+U_{n}\right) \label{Dn}%
\end{equation}
where, as in \cite{Yuki}, we introduced the logarithmic derivative
in $P$ of the general Liouville block
\begin{equation}
f_{P}\left(  x\right)  =i\frac d{dP}\log\mathcal{F}_{P}\left(
x\right)
\label{Flog}%
\end{equation}
The logarithms appearing in this derivative are due to a kind of
degeneracy which occur at equal external dimensions and leads to the
double poles in the integrand of (\ref{gG}). Apparently this effect
hides nothing conceptually new and there is no point to talk about
the Liouville field theory as of ``logarithmic
CFT''\footnote{Whatever this last term means.}. In the four point
function with different external dimensions there are no logarithms.
The constants $\mathcal{N}_{n}$ and $U_{n}$ in (\ref{Dn}) read
explicitly
\begin{align}
\mathcal{N}_{n}  & =\ \frac{\gamma^{(Q-4g)/b}(b^{2})\gamma(4gb-b^{2}+2nb^{2}%
)}{b^{2}\Gamma(2+b^{-2}-4gb^{-1})}\label{Nn}\\
& \ \ \ \ \times\prod_{k=0}^{n-1}\frac{\left(  4g-b^{-1}-b+kb\right)  ^{2}%
}{(4g-b^{-1}+(k+n-1)b)^{2}}\frac{\gamma^{2}(4gb+(k+n-1)b^{2})}{\gamma
^{4}(2gb+kb^{2})\gamma^{2}(1+(k+1)b^{2})}\nonumber
\end{align}
and
\begin{align}
U_{n}  & =2\upsilon_{b}(4g-Q+nb)-2\upsilon_{b}(4g-Q+2nb)+4\upsilon
_{b}(2g+nb)-2\upsilon_{b}(4g+2nb)\label{Un}\\
& \ \ \ \ -2\upsilon_{b}(b)-4(n+1)b\log b-4bC_{\text{E}}+2b\sum_{k=1}%
^{n}\left(  \psi(-kb^{2})+\psi(1+kb^{2})\right) \nonumber
\end{align}
In the last expression we have, as in ref. \cite{Yuki}, introduced
the notation
\begin{equation}
\upsilon_{b}(x)=\frac d{dx}\log\Upsilon_{b}(x)\label{ups}%
\end{equation}
for the logarithmic derivative of the $\Upsilon_{b}$-function. This
special function can be evaluated through the integral
representation
\begin{equation}
\upsilon_{b}(x)=\int_{0}^{\infty}dt\left(  \frac{\sinh(Q-2x)t}{\sinh
bt\sinh
b^{-1}t}-\frac{(Q-2x)}te^{-2t}\right) \label{iups}%
\end{equation}
convergent in the strip $0<\operatorname*{Re}x<Q$. Outside this
region one of the following relations, whichever more convenient,
can be used to render the argument to the strip of convergence
\begin{align}
\upsilon_{b}(x+b)-\upsilon_{b}(x)  & =b(-2\log b+\psi(bx)+\psi
(1-bx))\label{fups}\\
\upsilon_{b}(x+b^{-1})-\upsilon_{b}(x)  & =b^{-1}(2\log b+\psi(b^{-1}%
x)+\psi(1-b^{-1}x))\nonumber
\end{align}

\section{Elliptic modular parameter}

After all these preliminaries we can turn to the integral
(\ref{M4}). For the fixed area four point function it reads
\begin{equation}
\mathcal{I}_{4}=\left\langle \left(  \varphi e^{2g\phi}\right)  ^{4}%
\right\rangle
_{\text{LG}}^{(A=\pi)}=6\int_{\mathbf{G}}G_{\text{YL}}(x,\bar
x)g_{\text{L}}(x,\bar x)d^{2}x\label{M4A}%
\end{equation}
As in ref. \cite{GYL}, in eq.(\ref{M4A}) we used the symmetry of the
integral under the six element modular subgroup of projective group,
generated by the transformations $x\rightarrow1/x$ and
$x\rightarrow1-x$. This group divides
the complex plane of $x$ in $6$ regions, the fundamental region $\mathbf{G=}%
\{\operatorname*{Re}x<1/2;\;\left|  1-x\right|  <1\}$ and its $5$
images. The integral in (\ref{M4A}) is reduced to $\mathbf{G}$ while
the factor $6$ in front of the integral takes into account the
equivalent images.

It turns remarkably convenient to introduce the ``elliptic'' modular
parameter through the standard map
\begin{equation}
\tau=i\frac{K(1-x)}{K(x)}\label{tau}%
\end{equation}
where
\begin{equation}
K(x)=\frac12\int_{0}^{1}\frac{dt}{[t(1-t)(1-xt)]^{1/2}}\label{Kelliptic}%
\end{equation}
is the complete elliptic integral of the first kind. The integral
(\ref{M4A}) becomes
\begin{equation}
\mathcal{I}_{4}=6\pi^{2}\int_{\mathbf{F}}\left|  x(1-x)\theta_{3}%
^{4}(q)\right|  ^{2}G_{\text{YL}}(q,\bar q)g_{\text{L}}(q,\bar q)d^{2}%
\tau\label{EM4A}%
\end{equation}
where $\mathbf{F}$\textbf{\ }$=\left\{  \left|  \tau\right|
>1;\;\left| \operatorname*{Re}\tau\right|  <1/2\right\}  $\textbf{\
}is now the standard fundamental region of the modular group and
\begin{equation}
\theta_{3}(q)=\sum_{n=-\infty}^{\infty}q^{n^{2}}\label{T3}%
\end{equation}
is the usual $\theta$-series in
\begin{equation}
q=e^{i\pi\tau}\label{q}%
\end{equation}

There are two important advantages in the form (\ref{EM4A}). As it
has been argued in refs.\cite{block}, a general four point conformal
block admits a convenient recursive representation, which looks
particularly simple in terms of the elliptic parameter $q$ and the
so called ``elliptic'' four point block. Also, while the two blocks
(\ref{FIphi}) are known in closed form, the Liouville block
(\ref{Lblock}) at general $P$ is only computed as a power series in
$x$ or $q$. The recursive algorithm in the elliptic representation
gives directly the power series in $q$, which is argued to have much
better convergence then that in $x$. In particular, in the elliptic
fundamental
region $\max_{\mathbf{F}}\left|  q\right|  ^{2}=\exp(-\pi\sqrt{3}%
)=0.00433\ldots$ while in its $x$ image $\max_{\mathbf{G}}\left|
x\right| ^{2}=1$ near the cusps of the fundamental region, where the
convergence of the $x$-series is questionable.

Elliptic representations of the four point conformal blocks is
recapitulated in the Appendix. It is also shown there that the
products of the Yang-Lee and Liouville blocks read in the elliptic
parametrization as
\begin{equation}
x(1-x)\theta_{3}^{4}(q)\mathcal{F}_{\alpha}(x)\mathcal{F}_{P}(x)=\left(
16q\right)  ^{\Delta_{\alpha}+P^{2}+\lambda_{1,-1}^{2}}H_{\alpha}%
(q)H_{P}(q)H_{\text{sp}}(q)\label{HHH}%
\end{equation}
Here and below the index $\alpha=I,\varphi$ specifies the
intermediate representation in the Yang-Lee block, in particular
$\Delta_{I}=0$ and $\Delta_{\varphi}=-1/5$. The corresponding
elliptic blocks $H_{\alpha}(q)$ can be found as a series in $q$
either through the explicit expressions (\ref{FIphi})
\begin{align}
\mathcal{F}_{I}(x)  & =(16q)^{9/40}[x(1-x)]^{7/40}\theta_{3}^{1/2}%
(q)H_{I}(q)\label{FH}\\
\mathcal{F}_{\varphi}(x)  & =(16q)^{1/40}[x(1-x)]^{7/40}\theta_{3}%
^{1/2}(q)H_{\varphi}(q)\nonumber
\end{align}
or via the recursive relation (\ref{rec}). In both cases we have
\begin{align}
H_{I}(q)  & =1-\frac{21}{22}q^{2}-\frac{51}{88}q^{4}+\frac{1989}{5456}%
q^{6}-\frac{111489}{1789568}q^{8}+\frac{1612779}{3579136}q^{10}-\ldots
\label{Halpha}\\
H_{\varphi}(q)  & =1-\frac56q^{2}-\frac{91}{152}q^{4}+\frac{1967}{8816}%
q^{6}-\frac{8211}{70528}q^{8}+\frac{405647}{987392}q^{10}-\ldots\nonumber
\end{align}
The ``central charge deficit'' part
\begin{equation}
H_{\text{sp}}(q)=\left(  1-q^{2}-q^{4}+q^{10}+\ldots\right)  ^{-c_{\text{sp}%
}/2}\label{Hspq}%
\end{equation}
is carried out in the Appendix. Finally, the $q$-expansion of the
Liouville elliptic block is easily generated through recursive
relation (\ref{rec}). Although the calculation is straightforward,
the result is somewhat cumbersome at higher orders, so that here we
quote it explicitly only up to $O(q^{2})$ (the notation
$\lambda_{m,n}$ is explained in (\ref{lmn}))
\begin{equation}
H_{P}(q)=1+\frac{\left(  16p^{2}-b^{2}\right)  ^{2}q^{2}}{4\left(
1-b^{-4}\right)  \left(  P^{2}+\lambda_{1,2}^{2}\right)
}+\frac{\left( 16p^{2}-b^{-2}\right)  ^{2}q^{2}}{4\left(
1-b^{4}\right)  \left(
P^{2}+\lambda_{2,1}^{2}\right)  }+O(q^{4})\label{Hp2}%
\end{equation}
where $p=Q/2-g$. In numerical calculations below we used the
expansion of $H_{P}(q)$ up to the order $q^{8}$. In our symmetric
case (all external dimensions are equal) all the elliptic blocks are
series in $q^{2}$.

\section{Modular integral}

The Liouville correlation function can be separated into the
``integral part'' and, sometimes, a number of discrete terms
(\ref{DndP}). Consequently
\begin{equation}
\mathcal{I}_{4}=\sum_{n=0}^{N_{\text{d}}-1}\mathcal{I}_{\text{disc},\text{ }%
n}+\mathcal{I}_{\text{int}}\label{Idisc}%
\end{equation}
For numerical integration it is convenient to further separate each
term in two parts, corresponding to two matter blocks in
(\ref{G4YL})
\begin{align}
\mathcal{I}_{\text{int}}  & =J_{\text{int}}^{(I)}-\kappa^{2}J_{\text{int}%
}^{(\varphi)}\label{Jalpa}\\
\mathcal{I}_{\text{disc},\;n}  & =J_{\text{disc},\;n}^{(I)}-\kappa
^{2}J_{\text{disc},\;n}^{(\varphi)}\nonumber
\end{align}

Consider first the integral part
\begin{equation}
J_{\text{int}}^{(\alpha)}=\frac{3\pi
R_{g}}{\Gamma(4gb^{-1}-Qb^{-1})}\int
_{0}^{\infty}r_{g}(P)dP\int_{\mathbf{F}}\left|  (16q)^{P^{2}+Q^{2}%
/4-1+\Delta_{\alpha}}H_{P}^{(\alpha)}(q)\right|  ^{2}d^{2}\tau\label{Jint}%
\end{equation}
The product of the elliptic blocks
\begin{equation}
H_{P}^{(\alpha)}(q)=H_{\alpha}(q)H_{\text{sp}}(q)H_{P}(q)\label{Htot}%
\end{equation}
is developed in a power series in $q$
\begin{equation}
H_{\alpha}(q)H_{\text{sp}}(q)H_{P}(q)=\sum_{r=0}^{\infty}H_{r}^{(\alpha
)}(P)q^{r}\label{Hr}%
\end{equation}
so that the integrand in (\ref{Jint}) as a double power series in
$q$ and $\bar q$. In each term the integration in
$\tau_{2}=\operatorname*{Im}\tau$ can be carried out explicitly. The
result is in terms of the following function
\begin{equation}
\Phi(A,r.l)=\int_{\mathbf{F}}d^{2}\tau\left|  16q\right|
^{2A}q^{r}\bar
q^{l}=\frac{(16)^{2A}}{\pi(2A+r+l)}\int_{-1/2}^{1/2}\cos(\pi(r-l)x)e^{-\pi
\sqrt{1-x^{2}}\left(  2A+r+l\right)  }dx\label{Phi}%
\end{equation}
Notice, that if the integral is divergent at
$\tau_{2}\rightarrow\infty$, this reduction automatically takes care
of the divergency (in the sense of analytic continuation). We arrive
at the series
\begin{equation}
J_{\text{int}}^{(\alpha)}=\frac{3\pi R_{g}}{\Gamma(4gb^{-1}-Qb^{-1})}%
\sum_{L=0}^{\infty}A_{\text{int},\;L}^{(\alpha)}\label{JintL}%
\end{equation}
where in the last sum
\begin{equation}
A_{L}^{(\alpha)}=\int_{0}^{\infty}r_{g}(P)dP\sum_{k=0}^{L}H_{k}^{(\alpha
)}(P)H_{L-k}^{(\alpha)}(P)\Phi\left(  P^{2}+Q^{2}/4+\Delta_{\alpha
}-1,k,L-k\right) \label{Alevel}%
\end{equation}
In our symmetric case only even $L$ contribute. Each term in
(\ref{Alevel}) is suppressed by a factor $\max_{\mathbf{F}}\left|
q\right|  ^{2L}$ and in practice the series in $L$ converges very
fast. Below we found it sufficient to sum up to $L=6$ to reach the 8
-- 9 digit precision (see table \ref{Table1}).

The discrete terms, if any, are treated similarly. It is again
convenient to
single out the ``logarithmic'' part in each of the integrals $J_{\text{disc}%
,\;n}^{(\alpha)}$%
\begin{equation}
J_{\text{disc},\;n}^{(\alpha)}=6\pi^{2}\mathcal{N}_{n}\left(
L_{n}^{(\alpha
)}+K_{n}^{(\alpha)}\right) \label{JLK}%
\end{equation}
where ($P_{n}$ is from eq.(\ref{Pn}))
\begin{align}
L_{n}^{(\alpha)}  & =4\pi(Q/2-2g-nb)\int_{\mathbf{F}}\left|  (16q)^{P_{n}%
^{2}+\Delta_{\alpha}+Q^{2}/4-1}H_{P_{n}}^{(\alpha)}(q)\right|  ^{2}%
\operatorname*{Im}\tau\;d^{2}\tau\label{Logn}\\
K_{n}^{(\alpha)}  & =\int_{\mathbf{F}}\left|
(16q)^{P_{n}^{2}+\Delta_{\alpha
}+Q^{2}/4-1}H_{P_{n}}^{(\alpha)}(q)\right|  ^{2}\left(  \left.
-i\frac d{dP}\log H_{p}\bar H_{P}\right|  _{P=P_{n}}+S_{n}\right)
d^{2}\tau\nonumber
\end{align}
(only the Liouville elliptic blocks $H_{P}(q)$ depend on $P$ and
therefore appear in the logarithmic derivative) and
\begin{equation}
S_{n}=U_{n}-4(Q/2-2g-nb)\log16\label{Sn}%
\end{equation}
As in the case of the integral part, the expansion in $q$ and $\bar
q$ induces the ``level'' series
\begin{align}
L_{n}^{(\alpha)}  & =\sum_{L=0}^{\infty}B_{n,L}^{(\alpha)}\label{levs}\\
M_{n}^{(\alpha)}  & =\sum_{L=0}^{\infty}C_{n,L}^{(\alpha)}\nonumber
\end{align}
The integrals are then evaluated in the same way as (\ref{JintL})
\begin{align}
B_{n,L}^{(\alpha)}  & =4\pi(Q/2-2g-nb)\sum_{k=0}^{L}H_{k}^{(\alpha)}%
(P_{n})H_{L-k}^{(\alpha)}(P_{n})\Phi^{\prime}\left(  P_{n}^{2}+Q^{2}%
/4+\Delta_{\alpha}-1,k,L-k\right) \label{BCHH}\\
C_{n,L}^{(\alpha)}  & =\sum_{k=0}^{L}\left(  S_{n}H_{L-k}^{(\alpha)}%
(P_{n})-2{}H_{L-k}^{\prime\;(\alpha)}(P_{n})\right)  H_{k}^{(\alpha)}%
(P_{n})\Phi\left(  P_{n}^{2}+Q^{2}/4+\Delta_{\alpha}-1,k,L-k\right)
\nonumber
\end{align}
where
\begin{equation}
H_{k}^{\prime\;(\alpha)}(P)=i\frac d{dP}H_{k}^{(\alpha)}(P)\label{Hp}%
\end{equation}
A new function
\begin{align}
\  & \Phi^{\prime}(A,r.l)=\int d^{2}\tau\left|  16q\right|
^{2A}q^{r}\bar
q^{l}\operatorname*{Im}\tau\label{Phip}\\
\  & =\frac{(16)^{2A}}{(2A+r+l)^{2}\pi^{2}}\int_{-1/2}^{1/2}\cos
(\pi(r-l)x)e^{-\pi\sqrt{1-x^{2}}\left(  2A+r+l\right)  }\left(
1+\pi (2A+r+l)\sqrt{1-x^{2}}\right)  dx\nonumber
\end{align}
has been introduced to treat the logarithmic part. The ``level''
series also converges very fast (see next section).

\section{Numerical results}

In table \ref{Table1} we present the numerical values of the fourth
perturbative coefficient
\begin{equation}
z_{4}=\frac{\mathcal{I}_{4}}{24(2\pi)^{4}Z_{\text{L}}^{(A)}}\label{z4I4}%
\end{equation}
at different values of $b^{2}$ together with the second and third
ones (\ref{z2z3}) already evaluated in ref. \cite{GYL}. The forth
column contains the preliminary estimates $z_{4}^{\text{(est)}}$ of
$z_{4}$ on the basis of $z_{2}$, $z_{3}$ and analytic properties of
$z(h)$, as explained in \cite{GYL}. Also two ``exact'' values are
produced at the ``solvable'' points $b^{2}=0.4$ and $b^{2}=0.3$ (see
sect.8 for details).

\begin{center}%
\begin{table}[htb] \centering
\begin{tabular}
[c]{|llllll|}\hline \multicolumn{1}{|l|}{$b^{2}$} &
\multicolumn{1}{l|}{$z_{2}\times10^{2}$} &
\multicolumn{1}{l|}{$z_{3}\times10^{3}$} & \multicolumn{1}{l|}{$z_{4}%
\times10^{4}$} &
\multicolumn{1}{l|}{$z_{4}^{\text{(est)}}\times10^{4}$} &
$z_{4}^{\text{(exact)}}\times10^{4}$\\\hline
\multicolumn{1}{|l|}{$0.00$} & \multicolumn{1}{l|}{$-8.92857$} &
\multicolumn{1}{l|}{$22.9899$} & \multicolumn{1}{l|}{$-31.891804$} &
\multicolumn{1}{l|}{$-31.8938$} & \\\hline
\multicolumn{1}{|l|}{$0.01$} & \multicolumn{1}{l|}{$-8.83599$} &
\multicolumn{1}{l|}{$22.5977$} & \multicolumn{1}{l|}{$-31.114313$} &
\multicolumn{1}{l|}{$-31.1164$} & \\\hline
\multicolumn{1}{|l|}{$0.05$} & \multicolumn{1}{l|}{$-8.43801$} &
\multicolumn{1}{l|}{$20.9364$} & \multicolumn{1}{l|}{$-27.881220$} &
\multicolumn{1}{l|}{$-27.8839$} & \\\hline
\multicolumn{1}{|l|}{$0.10$} & \multicolumn{1}{l|}{$-7.86500$} &
\multicolumn{1}{l|}{$18.6240$} & \multicolumn{1}{l|}{$-23.557311$} &
\multicolumn{1}{l|}{$-23.5604$} & \\\hline
\multicolumn{1}{|l|}{$0.15$} & \multicolumn{1}{l|}{$-7.18331$} &
\multicolumn{1}{l|}{$16.0092$} & \multicolumn{1}{l|}{$-18.942809$} &
\multicolumn{1}{l|}{$-18.9461$} & \\\hline
\multicolumn{1}{|l|}{$0.20$} & \multicolumn{1}{l|}{$-6.35922$} &
\multicolumn{1}{l|}{$13.0616$} & \multicolumn{1}{l|}{$-14.131054$} &
\multicolumn{1}{l|}{$-14.1345$} & \\\hline
\multicolumn{1}{|l|}{$0.25$} & \multicolumn{1}{l|}{$-5.34942$} &
\multicolumn{1}{l|}{$9.78831$} & \multicolumn{1}{l|}{$-9.3340267$} &
\multicolumn{1}{l|}{$-9.33745$} & \\\hline
\multicolumn{1}{|l|}{$0.30$} & \multicolumn{1}{l|}{$-4.09998$} &
\multicolumn{1}{l|}{$6.28732$} & \multicolumn{1}{l|}{$-4.94949021$}
& \multicolumn{1}{l|}{$-4.95242$} & $-4.94949020548$\\\hline
\multicolumn{1}{|l|}{$0.35$} & \multicolumn{1}{l|}{$-2.55378$} &
\multicolumn{1}{l|}{$2.87061$} & \multicolumn{1}{l|}{$-1.61778168$}
& \multicolumn{1}{l|}{$-1.61953$} & \\\hline
\multicolumn{1}{|l|}{$0.40$} & \multicolumn{1}{l|}{$-0.71440$} &
\multicolumn{1}{l|}{$0.35675$} & \multicolumn{1}{l|}{$-0.085061507$}
& \multicolumn{1}{l|}{$-0.085303$} & $-0.08506150735$\\\hline
\end{tabular}
\caption{Numerical values for the second, third and forth order
perturbative coefficients in the fixed area scaling function $z(h)$.
In the forth coloumn we place the estimate of the four-point
coefficient from the sum rules, as explained in ref.\cite{GYL}.
Exact values, where available, are
presented for comparison. \label{Table1}}%
\end{table}%
\end{center}

At $b^{2}<b_{0}^{2}$ the fourth coefficient contains the
contributions from the integral part and discrete terms in
(\ref{DndP})
\begin{equation}
z_{4}=\sum_{n=0}^{N_{\text{d}}}z_{4}^{\text{(d)}}(n)+z_{4}^{\text{(int)}%
}\label{z4zdzint}%
\end{equation}
In table \ref{Table2} the structure of the four point integral as a
sum of discrete and integral contributions is illustrated
numerically. At small $b^{2} $ the integral part becomes negligible
while more and more discrete terms appear. In order, only few first
of these discrete terms really contribute at sufficiently small
$b^{2}$. E.g., at $b^{2}=0.01$ in fact there are as many as $49$
discrete terms and at $b^{2}=0$ (see next section) their number is
infinite. In the table we quote only those which count at the
precision level chosen (approximately $9$ decimal digits).

\begin{center}%
\begin{table}[htb] \centering
\begin{tabular}
[c]{|lllllll|}\hline \multicolumn{1}{|l|}{$b^{2}$} &
\multicolumn{1}{l|}{$0.00$} & \multicolumn{1}{l|}{$0.01$} &
\multicolumn{1}{l|}{$0.05$} & \multicolumn{1}{l|}{$0.10$} &
\multicolumn{1}{l|}{$0.15$} & $0.20$\\\hline
\multicolumn{1}{|l|}{$z_{4}^{\text{(int)}}\times10^{4}$} &
\multicolumn{1}{l|}{$0.00$} &
\multicolumn{1}{l|}{$8.6\times10^{-26}$} &
\multicolumn{1}{l|}{$0.006087$} & \multicolumn{1}{l|}{$1.7311$} &
\multicolumn{1}{l|}{$2.318$} & $0.799$\\\hline
\multicolumn{1}{|l|}{$z_{4}^{\text{(d)}}(0)\times10^{4}$} &
\multicolumn{1}{l|}{$-11.4450$} & \multicolumn{1}{l|}{$-10.5160$} &
\multicolumn{1}{l|}{$-6.88429 $} & \multicolumn{1}{l|}{$-4.187$} &
\multicolumn{1}{l|}{$-6.397$} & $-14.93$\\\hline
\multicolumn{1}{|l|}{$z_{4}^{\text{(d)}}(1)\times10^{4}$} &
\multicolumn{1}{l|}{$-15.7357$} & \multicolumn{1}{l|}{$-15.5090$} &
\multicolumn{1}{l|}{$-13.9653 $} & \multicolumn{1}{l|}{$-11.51$} &
\multicolumn{1}{l|}{$-14.864$} & \\\hline
\multicolumn{1}{|l|}{$z_{4}^{\text{(d)}}(2)\times10^{4}$} &
\multicolumn{1}{l|}{$-3.94363$} & \multicolumn{1}{l|}{$-4.16729$} &
\multicolumn{1}{l|}{$-4.97850 $} & \multicolumn{1}{l|}{$-5.770$} &
\multicolumn{1}{l|}{} & \\\hline
\multicolumn{1}{|l|}{$z_{4}^{\text{(d)}}(3)\times10^{4}$} &
\multicolumn{1}{l|}{$-0.662455$} & \multicolumn{1}{l|}{$-0.77641$} &
\multicolumn{1}{l|}{$-1.43314$} & \multicolumn{1}{l|}{$-3.829$} &
\multicolumn{1}{l|}{} & \\\hline
\multicolumn{1}{|l|}{$z_{4}^{\text{(d)}}(4)\times10^{4}$} &
\multicolumn{1}{l|}{$-0.092114$} & \multicolumn{1}{l|}{$-0.12415$} &
\multicolumn{1}{l|}{$-0.41407$} & \multicolumn{1}{l|}{} &
\multicolumn{1}{l|}{} & \\\hline
\multicolumn{1}{|l|}{$z_{4}^{\text{(d)}}(5)\times10^{4}$} &
\multicolumn{1}{l|}{$-0.011464$} & \multicolumn{1}{l|}{$-0.01846$} &
\multicolumn{1}{l|}{$-0.13010$} & \multicolumn{1}{l|}{} &
\multicolumn{1}{l|}{} & \\\hline
\multicolumn{1}{|l|}{$z_{4}^{\text{(d)}}(6)\times10^{4}$} &
\multicolumn{1}{l|}{$-0.001324$} & \multicolumn{1}{l|}{$-0.00265$} &
\multicolumn{1}{l|}{$-0.04666$} & \multicolumn{1}{l|}{} &
\multicolumn{1}{l|}{} & \\\hline
\multicolumn{1}{|l|}{$z_{4}^{\text{(d)}}(7)\times10^{4}$} &
\multicolumn{1}{l|}{$-1.44668\times10^{-4}$} &
\multicolumn{1}{l|}{$-3.76\times10^{-4}$} &
\multicolumn{1}{l|}{$-0.02083$} & \multicolumn{1}{l|}{} &
\multicolumn{1}{l|}{} & \\\hline
\multicolumn{1}{|l|}{$z_{4}^{\text{(d)}}(8)\times10^{4}$} &
\multicolumn{1}{l|}{$-1.51392\times10^{-5}$} &
\multicolumn{1}{l|}{$-5.34\times10^{-5}$} &
\multicolumn{1}{l|}{$-0.01443$} & \multicolumn{1}{l|}{} &
\multicolumn{1}{l|}{} & \\\hline
\multicolumn{1}{|l|}{$z_{4}^{\text{(d)}}(9)\times10^{4}$} &
\multicolumn{1}{l|}{$-1.52956\times10^{-6}$} &
\multicolumn{1}{l|}{$-7.65\times10^{-6}$} & \multicolumn{1}{l|}{} &
\multicolumn{1}{l|}{} & \multicolumn{1}{l|}{} & \\\hline
\multicolumn{1}{|l|}{$z_{4}^{\text{(d)}}(10)\times10^{4}$} &
\multicolumn{1}{l|}{$-1.50085\times10^{-7}$} &
\multicolumn{1}{l|}{$-1.11\times10^{-6}$} & \multicolumn{1}{l|}{} &
\multicolumn{1}{l|}{} & \multicolumn{1}{l|}{} & \\\hline
\multicolumn{1}{|l|}{$z_{4}^{\text{(d)}}(11)\times10^{4}$} &
\multicolumn{1}{l|}{$-1.43677\times10^{-8}$} &
\multicolumn{1}{l|}{$-1.66\times10^{-7}$} & \multicolumn{1}{l|}{} &
\multicolumn{1}{l|}{} & \multicolumn{1}{l|}{} & \\\hline
\multicolumn{1}{|l|}{$z_{4}^{\text{(d)}}(12)\times10^{4}$} &
\multicolumn{1}{l|}{$-1.34666\times10^{-9}$} &
\multicolumn{1}{l|}{$-2.53\times10^{-8}$} & \multicolumn{1}{l|}{} &
\multicolumn{1}{l|}{} & \multicolumn{1}{l|}{} & \\\hline
\end{tabular}
\caption{Relative contributions of the integral and discrete terms
in the sum
(\ref{z4zdzint}) at different values of $b^2$. \label{Table2}}%
\end{table}%
\end{center}

Finally, our results for $z_{4}$ are used in the analytic-numeric
procedure described in the first article \cite{GYL}. This allows to
correct the previous numerical results for the scaling function
$z(h)$. In particular we improve the numerical approximations for
the specific vacuum energy parameter
$f_{0}(b^{2})$ (see eq.(\ref{f0}) or \cite{GYL}). The new numbers $f_{0}%
^{(4)}$, which take into account the perturbative coefficients up to
$z_{4}$ are presented in table \ref{Table3} and compared with the
previous approximations as well as with the exact values
$f_{0}^{\text{(exact)}}$ where the last are available.

\begin{center}%
\begin{table}[htb] \centering
\begin{tabular}
[c]{|l|l|l|l|l|l|l|}\hline $b^{2}$ & $\rho$ & $\delta$ &
$f_{0}^{(2)}$ & $f_{0}^{(3)}$ & $f_{0}^{(4)}$ &
$f_{0}^{\text{(exact)}}$\\\hline $0.00$ & $0.833333$ & $0.111111$ &
$0.220407$ & $0.218156$ & $0.218036$ & $0.2179745$\\\hline $0.01$ &
$0.831646$ & $0.109935$ & $0.220318$ & $0.218091$ & $0.217959$ &
\\\hline
$0.05$ & $0.824462$ & $0.106179$ & $0.219719$ & $0.217523$ &
$0.217353$ &
\\\hline
$0.10$ & $0.814333$ & $0.103698$ & $0.218260$ & $0.215950$ &
$0.215742$ &
\\\hline
$0.15$ & $0.802587$ & $0.103880$ & $0.215654$ & $0.213053$ &
$0.212805$ &
\\\hline
$0.20$ & $0.788675$ & $0.107063$ & $0.211308$ & $0.208218$ &
$0.207920$ &
\\\hline
$0.25$ & $0.771700$ & $0.113797$ & $0.204224$ & $0.200408$ &
$0.200042$ &
\\\hline
$0.30$ & $0.75$ & $0.125$ & $0.192522$ & $0.187677$ & $0.187218$ &
$0.1870437 $\\\hline $0.35$ & $0.719788$ & $0.142247$ & $0.171896$ &
$0.165563$ & $0.164980$ &
\\\hline
$0.40$ & $0.666667$ & $0.166667$ & $0.125625$ & $0.116905$ &
$0.116151$ & $0.1158596$\\\hline
\end{tabular}
\caption {Specific energy $f_0$ determined with the use of first
two, three and four perturbative coefficients. When available, the
exact values are quoted
for comparison. \label{Table3}}%
\end{table}%
\end{center}

\section{Classical limit}

In ref. \cite{Classic} the classical limit of the Liouville four
point function has been considered. In particular, the ``symmetric''
function with four equal dimensions $\sigma$ admits the following
integral representation
\begin{equation}
g_{\text{cl}}(x,\bar x)=\pi^{3}\int_{\uparrow}\frac{(2s-1)ds}{2\pi i}%
\frac{\Gamma^{2}(2\sigma+s-1)\Gamma^{2}(2\sigma-s)\Gamma^{4}(s)}{\Gamma
^{4}(2\sigma)\Gamma^{2}(2s)}\mathcal{F}^{\text{(cl)}}(\sigma,s,y)\mathcal{F}%
^{\text{(cl)}}(\sigma,s,\bar y)\label{gcl}%
\end{equation}
The ``classical block'' $\mathcal{F}^{\text{(cl)}}(\sigma,s,y)$ is
expressed explicitly through the hypergeometric function
\begin{equation}
\mathcal{F}^{\text{(cl)}}(\sigma,s,y)=y^{s-2\sigma}{}_{2}F_{1}%
(s,s,2s,y)\label{cblock}%
\end{equation}
while the integration contour $\uparrow$ goes up along the imaginary
axis to the left from the poles of $\Gamma^{2}(2\sigma-s)$ and to
the right from all other singularities of the integrand. We are
going to argue that this expression is consistent with the classical
limit of the fixed area four-point function (\ref{gG}) provided we
identify $\Delta_{g}=\sigma$ and (\ref{gcl}) with the normalized
correlation function
\begin{equation}
\lim_{b^{2}\rightarrow0}\frac{g_{\text{L}}(x,\bar x)}{Z_{\text{L}}^{(A)}%
}=g_{\text{cl}}(x,\bar x)\label{climit}%
\end{equation}

To this order, let us evaluate the integral through the residues at
the infinite sequence of the ``right'' poles at $s=s_{n}$, where
\begin{equation}
s_{n}=2\sigma+n\label{sn}%
\end{equation}
This is a rightful procedure, as the large $s$ asymptotic of the
integrand shows. Thus
\begin{equation}
g_{\text{cl}}(x,\bar
x)=\sum_{n=0}^{\infty}D_{n}^{\text{(cl)}}(x,\bar
x)\label{Dsum}%
\end{equation}
where
\begin{equation}
D_{n}^{\text{(cl)}}(x,\bar x)=N_{n}^{\text{(cl)}}\left|  \mathcal{F}%
^{\text{(cl)}}(\sigma,s_{n},x)\right|  ^{2}\left(
2\operatorname*{Re}\left.
\frac d{ds}\log\mathcal{F}^{\text{(cl)}}(\sigma,s,x)\right|  _{s=s_{n}}%
+U_{n}^{\text{(cl)}}\right) \label{Dcl}%
\end{equation}
We introduced the notations
\begin{equation}
N_{n}^{\text{(cl)}}=\frac{\pi^{3}}{(4\sigma+2n-1)(n!)^{2}}\prod_{k=0}%
^{n-1}\frac{(2\sigma+k)^{4}}{(4\sigma+n+k-1)^{2}}\label{Ncl}%
\end{equation}
and
\begin{equation}
U_{n}^{\text{(cl)}}=4\psi(4\sigma+2n)-2\psi(4\sigma+n-1)-4\psi(2\sigma
+n)+2\psi(1+n)-\frac2{4\sigma+2n-1}\label{Ucl}%
\end{equation}
Comparing (\ref{Ncl}) and (\ref{Ucl}) with (\ref{Nn}) and (\ref{Un})
respectively, it is easy to see that
\begin{align}
\lim_{b\rightarrow0}\frac{\mathcal{N}_{n}}{bZ_{\text{L}}}  & =N_{n}%
^{\text{(cl)}}\label{Climit}\\
\lim_{b\rightarrow0}bU_{n}  & =U_{n}^{\text{(cl)}}\nonumber
\end{align}
Also, it is well known (see e.g. \cite{block}) that
\begin{equation}
\lim_{b^{2}\rightarrow0}\mathcal{F}_{P}\left(  \left.
\begin{array}
[c]{cc}%
\sigma & \sigma\\
\sigma & \sigma
\end{array}
\right|  x\right)  =\mathcal{F}^{\text{(cl)}}(\sigma,s,x)\label{Blimit}%
\end{equation}
provided $\sigma$ and $s=P^{2}+Q^{2/4}$ are kept finite in the
limit. Thus, the classical limit of the normalized discrete term
(\ref{Dn}) coincides with (\ref{Dcl})
\begin{equation}
\lim_{b^{2}\rightarrow0}\frac{D_{n}(x,\bar x)}{Z_{\text{L}}^{(A)}}%
=D_{n}^{\text{(cl)}}(x,\bar x)\label{Dlimit}%
\end{equation}
In the classical limit the integral in eq.(\ref{DndP}) is saturated
by the infinite sequence of discrete terms, the integral one
vanishing. This proves (\ref{climit}).

The classical block can be rendered to the form (cp. eq.(\ref{FMFL})
in the Appendix)
\begin{equation}
F_{\alpha}\mathcal{F}^{\text{(cl)}}(\sigma,s,x)=\frac{(16q)^{\Delta_{\alpha
}+s-1}H_{\alpha}(q)H^{\text{(cl)}}(q)}{x(1-x)\theta_{3}^{4}}\label{FclFa}%
\end{equation}
In the general expression (\ref{FMFL}) only the product $H^{\text{(cl)}%
}(q)=H_{\text{sp}}H_{\text{L}}$ allows the classical limit. The
classical elliptic block reads
\begin{equation}
H^{\text{(cl)}}(q)=\eta^{c_{\text{M}}/2-25/2}(q^{2})\theta_{0}^{16\sigma
}(q^{2})h_{s}(q)\label{Hcl}%
\end{equation}
In this expression
\begin{align}
h_{s}(q)  & =\left(  \frac x{16q}\right)  ^{s}{}_{2}F_{1}(s,s,2s,x)\label{hsF}%
\\
\  & =1-\frac{4s(2s-3)}{2s+1}q^{2}+\frac{2\,s\,\left(  8\,s^{2}%
-14\,s+9\right)  }{2\,s+3}q^{4}+\ldots\nonumber
\end{align}
while
\begin{equation}
\eta(q^{2})=1-q^{2}-q^{4}+q^{10}+\ldots\label{eta}%
\end{equation}
is the standard Dedekind product (\ref{ded}) and
\begin{equation}
\theta_{0}(q^{2})=1-2q^{2}+2q^{8}-2q^{18}+\ldots=\sum_{n=-\infty}^{\infty
}(-)^{n}q^{2n^{2}}\label{T0}%
\end{equation}
the usual theta series. It is also straightforward to verify
directly that
$H^{\text{(cl)}}(q)$ is the limit of $H_{\text{sp}}H_{P}$ as $b^{2}%
\rightarrow0$ and $P^{2}\rightarrow s-Q^{2}/4$.

For our particular application in the GYL model the classical
elliptic block is evaluated through the explicit formula
\begin{equation}
H^{\text{(cl)}}(q)=\eta^{-147/10}(q^{2})\theta_{0}^{96/5}(q^{2})h_{s}%
(q)\label{hs}%
\end{equation}
while the matter elliptic blocks $H_{\alpha}$ remains the same as in
eq.(\ref{Halpha}). As usual, we are going to use the $q$-expansions.
Denote
\begin{align}
H_{\alpha}(q)H^{\text{(cl)}}(q)  & =\sum_{k=0}^{\infty}h_{k}^{(\alpha)}%
q^{k}\label{hk}\\
H_{\alpha}(q)\frac d{ds}H^{\text{(cl)}}(q)  & =\sum_{k=0}^{\infty}%
h_{k}^{\prime\;(\alpha)}q^{k}\nonumber
\end{align}
In the classical case the integral (\ref{M4A}) is the sum of the
infinite number of discrete terms
\begin{equation}
\mathcal{I}_{4}^{\text{(cl)}}=\sum_{n=0}^{\infty}I_{n}^{\text{(cl)}%
}\label{Icl}%
\end{equation}
where
\begin{equation}
I_{n}^{\text{(cl)}}=6\int_{\mathbf{G}}D_{n}^{\text{(cl)}}(x,\bar
x)G_{\text{YL}}(x,\bar
x)d^{2}x=j_{n}^{(I)}-\kappa^{2}j_{n}^{(\varphi
)}\label{Incl}%
\end{equation}
Now, as in the general case, we single out the ``logarithmic''
integral
\begin{equation}
j_{n}^{(\alpha)}=6\pi^{2}N_{n}^{\text{(cl)}}\left(  L_{n}^{(\alpha)}%
+M_{n}^{(\alpha)}+K_{n}^{(\alpha)}\right) \label{jn}%
\end{equation}
where,
\begin{align}
L_{n}^{(\alpha)}  & =2\pi\int_{\mathbf{F}}\left|
(16q)^{\Delta_{\alpha
}+2\sigma+n-1}h_{\alpha}(q)\right|  ^{2}\operatorname*{Im}\tau\,d^{2}%
\tau\nonumber\\
M_{n}^{(\alpha)}  & =\int_{\mathbf{F}}\left|  (16q)^{\Delta_{\alpha}%
+2\sigma+n-1}h_{\alpha}(q)\right|  ^{2}\left(  U_{n}^{\text{(cl)}}%
-2\log16\right)  \,d^{2}\tau\label{Mnint}\\
K_{n}^{(\alpha)}  & =-\int_{\mathbf{F}}\left|  (16q)^{\Delta_{\alpha}%
+2\sigma+n-1}\right|  ^{2}\frac d{ds}\left(
h_{\alpha}(q)h_{\alpha}(\bar q)\right)  d^{2}\tau\nonumber
\end{align}
Then, each term is obtained as a ``level by level'' sum in $L$
through the double $q$ and $\bar q$-expansions.
\begin{align}
L_{n}^{(\alpha)}  &
=2\pi\sum_{L=0}^{\infty}\sum_{k=0}^{L}h_{k}^{(\alpha
)}(s_{n})h_{L-k}^{(\alpha)}(s_{n})\Phi^{\prime}(\Delta_{\alpha}+s_{n}%
-1,k,L-k)\nonumber\\
M_{n}^{(\alpha)}  & =\left(  U_{n}^{\text{(cl)}}-2\log16\right)
\sum
_{L=0}^{\infty}\sum_{k=0}^{L}h_{k}^{(\alpha)}(s_{n})h_{L-k}^{(\alpha)}%
(s_{n})\Phi(\Delta_{\alpha}+s_{n}-1,k,L-k)\label{Ln}\\
K_{n}^{(\alpha)}  & =-2\sum_{L=0}^{\infty}\sum_{k=0}^{L}h_{k}^{(\alpha)}%
(s_{n})h_{L-k}^{\prime\;(\alpha)}(s_{n})\Phi(\Delta_{\alpha}+s_{n}%
-1,k,L-k)\nonumber
\end{align}
Every component is straightforwardly evaluated with the use of the
integrals (\ref{Phi}) and (\ref{Phip}). In practical calculations
(given the required precision of $9$ digits) we found it sufficient
to sum up to $L=10$ in eq.(\ref{Ln}).

Through all these calculations we arrive at the contribution of
$n$-th discrete term to the fixed area perturbative coefficient
\begin{equation}
z_{4}^{\text{(cl)}}(n)=\frac{I_{n}^{\text{(cl)}}}{24(2\pi)^{4}}\label{z4ncl}%
\end{equation}
Several first contributions are presented in table \ref{Table2} to
manifest the convergence. The discrete terms sum up to the number
\begin{equation}
z_{4}^{\text{(cl)}}=\sum_{n=0}^{\infty}z_{4}^{\text{(cl)}}%
(n)=-31.8918039\times10^{-4}\label{zclnum}%
\end{equation}
already obtained earlier \cite{Classic} by means of a different
numerical approach. Notice, that we had to take into account as many
as $10$ discrete terms to achieve this $9$ digit precision quoted.

\section{Integrable points $b^{2}=0.3$ and $b^{2}=0.4$}

In this section we comment about the two exactly solvable points
$b^{2}=0.4$ and $b^{2}=0.3$ in the family of GYL models. The
solution at the ``pure Yang-Lee'' point $b^{2}=2/5$ through the
matrix model approach has been already discussed in the previous
article \cite{GYL}. Here we recapitulate the essence very briefly.
This case of the ``minimal gravity'' is related to the flow from the
tricritical to the critical points in the generic one matrix
model \cite{Staudacher}. The corresponding scaling function $Z_{\text{YL}%
}(x,t)$, which is interpreted (up to an overall scale) as the
spherical partition function of the continuous gravity, is
determined explicitly as
\begin{equation}
\frac{\partial^{2}}{\partial x^{2}}Z_{\text{YL}}(x,t)=u(x,t)\label{Zt}%
\end{equation}
through a solution $u(x,t)$ of the following simple algebraic
equation
\begin{equation}
x=u^{3}-tu\label{u3}%
\end{equation}
Parameters $t$ and $x$ are, again up to some normalization
constants, the cosmological constant and $\varphi$-perturbation
coupling respectively. Comparing the series expansion generated by
(\ref{Zt}) and (\ref{u3})
\begin{equation}
Z_{\text{YL}}(x,t)=Z_{\text{YL}}(0,t)\left(  1-\frac{105}{16}\left(
\frac x{t^{3/2}}\right)  +\frac{105}8\left(  \frac x{t^{3/2}}\right)
^{2}-\frac
{35}{16}\left(  \frac x{t^{3/2}}\right)  ^{3}+\ldots\right) \label{Ztexp}%
\end{equation}
with the perturbative coefficients evaluated in the field theoretic
approach, it is easy to relate
\begin{equation}
\frac x{t^{3/2}}=\frac{\lambda l_{\text{YL}}}{(\pi\mu)^{3/2}}\label{xtmu}%
\end{equation}
where
\begin{equation}
l_{\text{YL}}=\frac{\gamma^{1/2}(4/5)}{4\gamma(2/5)}=0.0845223\ldots
\label{lYL}%
\end{equation}
Then, the fixed area scaling function (\ref{zh}) at $b^{2}=0.4$
reads explicitly
\begin{equation}
z(h)=-2\sqrt{\pi}\sum_{n=0}^{\infty}\frac{(l_{\text{YL}}h)^{n}}{n!\Gamma
(n/2-1/2)}\label{zh04}%
\end{equation}
In particular
\begin{align}
z_{2} &  =-l_{\text{YL}}^{2}=-0.00714401\nonumber\\
z_{3} &  =\frac{\sqrt{\pi}}3l_{\text{YL}}^{3}=0.000356752\label{zYLnum}\\
z_{4} &  =-\frac16l_{\text{YL}}^{4}=-8.50615\ldots\times10^{-6}\nonumber\\
&  \ldots\nonumber
\end{align}
The last number and
\begin{equation}
f_{0}(0.4)=\frac3\pi\left(  \frac{l_{\text{YL}}}2\right)  ^{2/3}%
=0.11585962159187\label{mc25}%
\end{equation}
are quoted in tables \ref{Table1} and \ref{Table2} as the
corresponding exact values for this point.

In ref. \cite{GYL} it has been argued that at $b^{2}=0.3$ our GYL
model is also a version of minimal gravity. It arises as a
particular perturbation of the minimal $\mathcal{M}_{3/10}$ model
coupled to the Liouville gravity. Therefore there are serious
reasons to believe that at this point the model is again exactly
solvable. Unfortunately yet no exact solution in the framework of
the matrix model approach is known. However, it is very natural to
expect the existence of a closed analytic expression, e.g., for the
scaling function (\ref{zh}). We conjecture the following explicit
form, apparently motivated by more general structures discovered in
\cite{Kostov} (see also \cite{threep})
\begin{equation}
z_{\text{3/10}}(h)=\Gamma(-1/3)\sum_{n=0}^{\infty}\frac{(l_{\text{eg}}h)^{n}%
}{n!\Gamma(n/3-1/3)}\label{z310}%
\end{equation}
where the scale factor $l_{\text{eg}}$ is easily figured out through
the comparison with the perturbative coefficients of sect.1
\begin{equation}
l_{\text{eg}}=\frac{\gamma(1/3)}{2^{1/2}3\gamma^{5/6}(3/10)}=0.23254\ldots
\label{leg}%
\end{equation}
This explicit expression gives rise to the following values of the
first perturbative coefficients
\begin{align}
z_{2}  & =-0.0409998231725\nonumber\\
z_{3}  & =0.00628731873887\label{z2345}\\
z_{4}  & =-0.000494949020548\nonumber\\
z_{5}  & =0.0000257778956523\nonumber
\end{align}
We consider the comparison of these numbers with those in the
corresponding row of the table \ref{Table1} as a convincing support
in favor of our conjecture. Analyzing the asymptotic of (\ref{z310})
one finds, in addition, that at $b^{2}=0.3$
\begin{equation}
f_{0}=\frac4\pi\left(  \frac{l_{\text{eg}}}3\right)
^{3/4}=0.187044\ldots
\label{f310}%
\end{equation}
This number is produced in table \ref{Table2} as the corresponding
``exact'' value.

\begin{figure}
[tbh]
\begin{center}
\includegraphics[
height=3.40in, width=3.00in
]%
{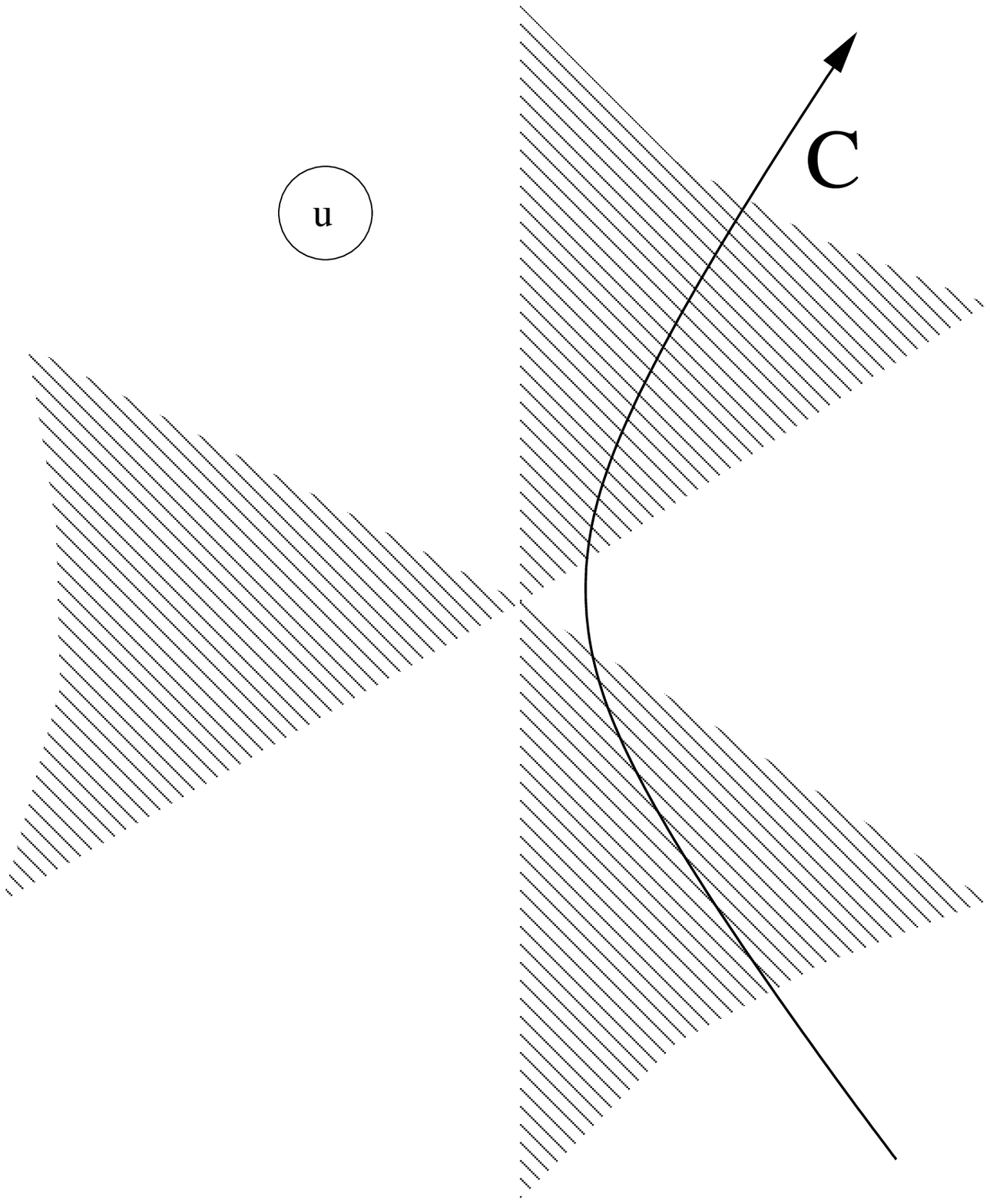}%
\caption{Integration contour in the representation (\ref{Ay}).
Dashed are the
wedges where the integrand decreases at large $\left|  u\right|  $.}%
\label{AiryC}%
\end{center}
\end{figure}

It seems also suggestive that the conjectured fixed area scaling
function (\ref{z310}) follows from the ``matrix like'' algebraic
equation
\begin{equation}
u^{4}-xu=-t\label{u4}%
\end{equation}
for the third derivative
\begin{equation}
u=\frac{\partial^{3}}{\partial t^{3}}Z_{\text{3/10}}(x,t)\label{Zttt}%
\end{equation}
of the ``grand'' partition function
\begin{align}
& Z_{\text{3/10}}(x,t)=-\frac{x^{13/3}}3\sum_{n=0}^{\infty}\frac
{\Gamma(4n/3-13/3)(tx^{-4/3})^{n}}{n!\Gamma(n/3-1/3)}\label{Z310}\\
& =-\frac{x^{13/3}}3\left(  \frac{81}{3640}-\frac{tx^{-4/3}}{24\,}%
+\frac{9\,(tx^{-4/3})^{2}}{20\,}-\frac{(tx^{-4/3})^{3}}{2\,}+\frac
{(tx^{-4/3})^{4}}{24}+\frac{(tx^{-4/3})^{5}}{90}+O(t^{6})\right)
\nonumber
\end{align}
Here the cosmological constant $x$ related to $\mu$ through
\begin{equation}
\frac{\lambda l_{\text{eg}}}{(\pi\mu)^{4/3}}=\frac t{x^{4/3}}\label{lmutx}%
\end{equation}

Finally, let us mention a convenient integral representation for the
scaling function (\ref{z310})
\begin{equation}
z_{\text{3/10}}(h)=3\int_{C}\exp\left(
u^{3}+hl_{\text{eg}}u^{-1}\right)
u^{3}du\label{Ay}%
\end{equation}
where the integration contour $C$ goes as it is shown in fig.\ref{AiryC}%

\textbf{Acknowledgments}

Moral support by Galchenok was always extremely important for me,
particularly while pursuing this little popular development. Hence
her encouragement is of unique value. The same concerns several very
interesting discussions with A.Neveu. I thank also A.Belavin for
interest to the work and sharing my conviction in its significance.
Most tedious calculations were carried out while visiting the
Bogolubov Laboratory of Theoretical Physics in Dubna and the Theory
Division of ITEP in December 2005 -- January 2006. I acknowledge the
hospitality, stimulating scientific air of these groups, and
personally A.Isaev and A.Losev. This travel was supported
(partially) by CNRS UMR 5825. I was also sponsored by the INTAS
project under grant \#INTAS-OPEN-03-51-3350 and by the European
Committee under contract EUCLID HRPN-CT-2002-00325.

\appendix

\section{Elliptic four point block}

Here we summarize the results of refs.\cite{block}. Consider a
conformal theory with central charge $c$. It will prove convenient
to introduce the
``Liouville like'' parameterization in terms of $b$%
\begin{equation}
\frac{c-1}6=\left(  b^{-1}+b\right)  ^{2}\label{cb}%
\end{equation}
This is not a restriction for the value of $c$ since we allow $b$ to
be complex if needed. It is also convenient to introduce the
notation
\begin{equation}
\lambda_{m,n}=\frac{mb^{-1}+nb}2\label{lmn}%
\end{equation}
Let $\Delta_{i}$, $i=1,2,3,4$ and $\Delta$ be the external and
intermediate dimensions in the four point block \cite{BPZ}, as it is
illustrated in the picture below
\begin{equation}
\mathcal{F}_{\Delta}\left(  \left.
\begin{array}
[c]{cc}%
\Delta_{1} & \Delta_{3}\\
\Delta_{2} & \Delta_{4}%
\end{array}
\right|  x\right)  =%
\raisebox{-0.4696in}{\includegraphics[ trim=0.000000in 0.000000in
0.001589in -0.000624in, height=1.0395in, width=2.2693in
]%
{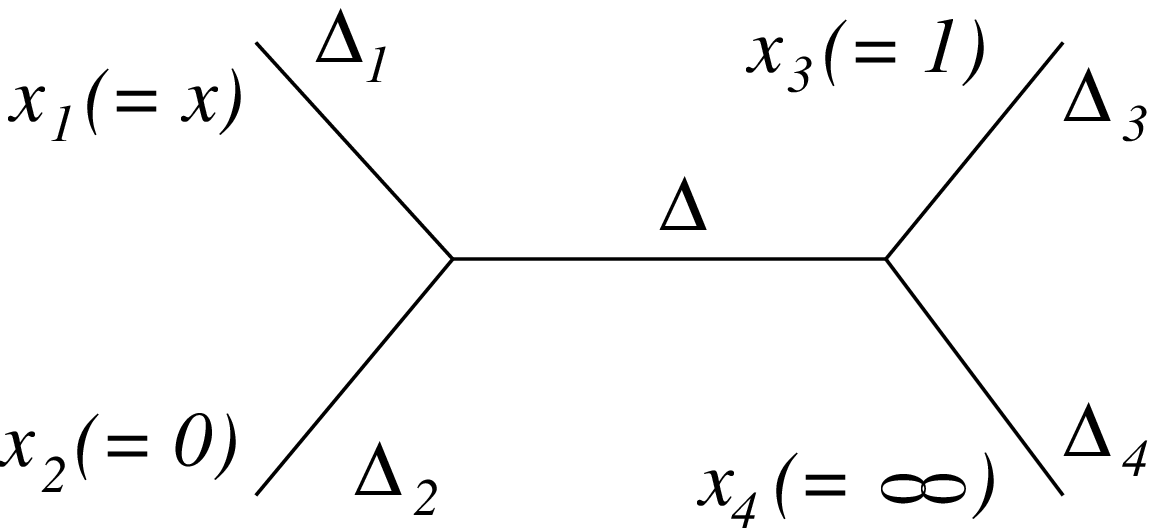}%
}%
\label{Block}%
\end{equation}
According to \cite{block} it can be written as
\begin{align}
\  & \mathcal{F}_{\Delta}\left(  \left.
\begin{array}
[c]{cc}%
\Delta_{1} & \Delta_{3}\\
\Delta_{2} & \Delta_{4}%
\end{array}
\right|  x\right)  =\label{Hblock}\\
& \ \ \left(  16q\right)  ^{\Delta-(c-1)/24}x^{(c-1)/24-\Delta_{1}-\Delta_{2}%
}(1-x)^{(c-1)/24-\Delta_{1}-\Delta_{3}}\theta_{3}^{(c-1)/2-4\sum_{i=1}%
^{4}\Delta_{i}}(q)H_{\Delta}\left(  q\right) \nonumber
\end{align}
where $q$ is related to $x$ as in eqs. (\ref{tau}), (\ref{q}) and
$\theta _{3}(q)$ is defined in (\ref{T3}). In the function
\begin{equation}
H_{\Delta}(q)=H_{\Delta}\left(  \left.
\begin{array}
[c]{cc}%
\lambda_{1} & \lambda_{3}\\
\lambda_{2} & \lambda_{4}%
\end{array}
\right|  q\right) \label{HDq}%
\end{equation}
we suppressed the dependence on the external dimensions, which are
conveniently parameterized in terms of $\lambda_{i}$, $i=1,2,3,4$ as
\begin{equation}
\Delta_{i}=\frac{c-1}{24}+\lambda_{i}^{2}\label{Dl}%
\end{equation}
The $H$-function (the elliptic four point block) is a power series in $q$%
\begin{equation}
H_{\Delta}\left(  \left.
\begin{array}
[c]{cc}%
\lambda_{1} & \lambda_{3}\\
\lambda_{2} & \lambda_{4}%
\end{array}
\right|  q\right)  =\sum_{L=0}^{\infty}H_{L}\left(  \left.
\begin{array}
[c]{cc}%
\lambda_{1} & \lambda_{3}\\
\lambda_{2} & \lambda_{4}%
\end{array}
\right|  \Delta\right)  q^{L}\label{HL}%
\end{equation}
which is believed to converge at $\left|  q\right|  <1$. It can be
effectively calculated through the recursive relation \cite{block}
\begin{equation}
H_{\Delta}\left(  \left.
\begin{array}
[c]{cc}%
\lambda_{1} & \lambda_{3}\\
\lambda_{2} & \lambda_{4}%
\end{array}
\right|  q\right)  =1+\sum_{(m,n)}\frac{q^{mn}}{\Delta-\Delta_{m,n}}%
R_{m,n}\left(
\begin{array}
[c]{cc}%
\lambda_{1} & \lambda_{3}\\
\lambda_{2} & \lambda_{4}%
\end{array}
\right)  H_{\Delta_{m,n}+mn}\left(  \left.
\begin{array}
[c]{cc}%
\lambda_{1} & \lambda_{3}\\
\lambda_{2} & \lambda_{4}%
\end{array}
\right|  q\right) \label{rec}%
\end{equation}
Here the sum is over all pairs $(m,n)$ of positive integers and
\begin{equation}
\Delta_{m,n}=\lambda_{1,1}^{2}-\lambda_{m,n}^{2}\label{Dmn}%
\end{equation}
are the dimensions of degenerate representations of the Virasoro
algebra with the central charge $c$. The multipliers $R_{m,n}$ in
(\ref{rec}) read explicitly
\begin{equation}
R_{m,n}(\lambda_{i})=2\ \frac{\prod_{r,s}(\lambda_{1}+\lambda_{2}%
-\lambda_{r,s})(\lambda_{1}-\lambda_{2}-\lambda_{r,s})(\lambda_{3}+\lambda
_{4}-\lambda_{r,s})(\lambda_{3}-\lambda_{4}-\lambda_{r,s})}{\prod
_{k,l}^{\prime}\lambda_{k,l}}\label{Rmn}%
\end{equation}
The products in (\ref{Rmn}) are over the following sets of integers
$(r,s)$ and $(k,l)$
\begin{align}
r &  =-m+1,-m+3,\ldots,m-3,m-1\label{rs}\\
s &  =-n+1,-n+3,\ldots,n-3,n-1\nonumber
\end{align}
and
\begin{align}
k &  =-m+1,-m+2,\ldots,m-1,m\label{kl}\\
l &  =-n+1,-n+2,\ldots,n-1,n\nonumber
\end{align}
while the prime sign near the last product symbol
$\prod_{k,l}^{\prime}$ means that the two pairs $(k,l)=(0,0)$ and
$(m,n)$ are missing.

Relation (\ref{rec}) leads, in particular, to a recursive algorithm
for the coefficients in the ``level'' expansion (\ref{HL})
\begin{align}
H_{0}(\Delta)  & =1\label{recL}\\
H_{L}(\Delta)  & =\sum_{mn<L}\frac{R_{m,n}}{\Delta-\Delta_{m,n}}%
H_{L-mn}(\Delta_{m,n}+mn)\nonumber
\end{align}
where we have again suppressed the dependence on $\lambda_{i}$.

Now for the purposes of quantum gravity we want to combine two
blocks of different conformal field theories, conventionally be the
``matter'' and the ``Liouville'' one, with central charges
respectively $c_{\text{M}}$ and $c_{\text{L}}$. We do not
necessarily require the ``complementarity'' of these quantities,
introducing the ``spectator'' central charge
\begin{equation}
c_{\text{sp}}=26-c_{\text{M}}-c_{\text{L}}\label{csp}%
\end{equation}
to take care of the deficit. On the contrary we do require the
complementarity of the ``matter'' and ``Liouville'' external
dimensions $\Delta_{i}$ and $\tilde\Delta_{i}$
\begin{equation}
\Delta_{i}+\tilde\Delta_{i}=1\label{DMDL}%
\end{equation}
for $i=1,2,3,4$. The ``matter'' and ``Liouville'' blocks are
combined to
\begin{equation}
\ \mathcal{F}_{\Delta}^{\text{(M)}}\left(  \left.
\begin{array}
[c]{cc}%
\Delta_{1} & \Delta_{3}\\
\Delta_{2} & \Delta_{4}%
\end{array}
\right|  x\right)  \mathcal{F}_{\tilde\Delta}^{\text{(L)}}\left(
\left.
\begin{array}
[c]{cc}%
\tilde\Delta_{1} & \tilde\Delta_{3}\\
\tilde\Delta_{2} & \tilde\Delta_{4}%
\end{array}
\right|  x\right)  =\ \left(  16q\right)  ^{\Delta+\tilde\Delta-1}%
\frac{H_{\Delta}^{\text{(M)}}(q)H_{\tilde\Delta}^{\text{(L)}}(q)H_{\text{sp}%
}(q)}{x(1-x)\theta_{3}^{4}(q)}\label{FMFL}%
\end{equation}
where we conventionally denoted
\begin{equation}
H_{\text{sp}}(q)=\eta^{-c_{\text{sp}}/2}(q^{2})\label{Hsp}%
\end{equation}
with
\begin{equation}
\eta(q^{2})=\prod_{k=1}^{\infty}\left(  1-q^{2k}\right)
=\sum_{n=-\infty
}^{\infty}(-)^{n}q^{n(3n+1)}\label{ded}%
\end{equation}
the Dedekind function.

Notice, that we \emph{do not }demand the intermediate dimensions
$\Delta$ and $\tilde\Delta$ to be complementary. To avoid
misunderstanding, let us stress that $H_{\text{sp}}(q)$ is simply a
convenient notation and hardly can be interpreted as a
``contribution of the spectator matter'' to the block and the four
point function. Some additional simplifications of the product of
the elliptic blocks, which occur if the ``matter'' and ``Liouville''
CFT's are indeed complementary (i.e., $c_{\text{sp}}=0$) will be
discussed in \cite{fp}.


\begin{thebibliography}{99}
\bibitem{KAT}K.A.Ter-Martirosyan. Lectures on theoretical physics. MFTI, 1974--1976.

\bibitem {Classic}Al.Zamolodchikov. Scaling Lee-Yang model on a sphere. I:
Partition function. JHEP 0207 (2002) 029; hep-th/0109078.

\bibitem {GYL}Al.Zamolodchikov. Perturbed conformal field theory on
fluctuating sphere. Contribution to the Balkan Workshop BW2003,
``Mathematical, Theoretical and Phenomenological Challenges Beyond
Standard Model'', 29 August--02 September 2003, Vrnjancka Banja,
Serbia, hep-th/0508044.

\bibitem {Yuki}Y.Ishimoto and Al.Zamolodchikov. Massive Majorana fermion
coupled to 2D gravity and random lattice Ising model.
hep-th/0510214, pp. 47--72.

\bibitem {BPZ}A.Belavin, A.Polyakov and A.Zamolodchikov. Infinity conformal
symmetry in two-dimensional quantum field theory. Nucl.Phys., B241
(1984) 333.

\bibitem {Cardy}J.Cardy. Conformal invariance and the Yang-Lee edge
singularity in two-dimensions. Phys.Rev.Lett., 54 (1985) 1354

\bibitem {AAl}A.Zamolodchikov and Al.Zamolodchikov. Structure constants and
conformal bootstrap in Liouville field theory. Nucl.\ Phys.\ B477
(1996) 577. hep-th/9506136.

\bibitem {threep}Al.Zamolodchikov. On the three-point function in minimal
Liouville gravity. hep-th/0505063.

\bibitem {KPZ}V.Knizhnik, A.Polyakov and A.Zamolodchikov. Fractal structure of
2--D quantum gravity. Mod.Phys.Lett., A3 (1988) 819.

\bibitem {block}Al.Zamolodchikov. Conformal symmetry in two-dimensions: an
explicit recurrence formula for the conformal partial wave
amplitude. Commun.Math.Phys.,\ 96 (1984) 419;

Conformal symmetry in two-dimensional space: recursion
representation of conformal block. Theor.Math.Phys., 73 (1987)
1088--1093.

\bibitem {Polyakov}A.Polyakov. Quantum geometry of bosonic strings.
Phys.Lett., B103 (1981) 207.

\bibitem {fp}Al.Zamolodchikov. Four point function in Liouville gravity.
Analytics and numerics. In preparation.

\bibitem {Staudacher}M.Staudacher. The Yang-Lee edge singularity on a
dynamical planar random surface. Nucl.Phys.,\ B336 (1990) 349.

\bibitem {Kostov}I.Kostov. Thermal flow in the gravitational O(n) model. Talk
delivered at the Fourth International Symposium ``Quantum Theory and
Symmetries'', Varna, Bulgaria, 15-21 August 2005. hep-th/0602075.
\end{thebibliography}
\end{document}